\documentclass[12pt]{article}
\usepackage{graphicx}

\begin{document}

\begin{center}

\Large

\textbf{Solar flare-related eruptions followed by long-lasting
occultation of the emission in the He~II 304~\AA\ line and in
microwaves}

\end{center}

\centerline{V.~V.~Grechnev$^1$, I.~V.~Kuzmenko$^2$,
I.~M.~Chertok$^3$, A.~M.~Uralov$^1$}

\begin{center}
% email{grechnev@iszf.irk.ru}
$^1$Institute of Solar-Terrestrial Physics SB RAS, Irkutsk, email: grechnev@iszf.irk.ru \\
$^2$Ussuriysk Astrophysical Observatory,
Primorsky Krai, Ussuriysk, email: kuzmenko\_irina@mail.ru \\
$^1$Pushkov Institute of Terrestrial Magnetism, Ionosphere and
Radio Wave Propagation (IZMIRAN), Troitsk, Moscow Region, email:
ichertok@izmiran.ru
\end{center}

%\date{\today}
\date{}

\begin{abstract}
Plasma with a temperature close to the chromospheric one is
ejected in solar eruptions. Such plasma can occult some part of
emission of compact sources in active regions as well as quiet
solar areas. Absorption phenomena can be observed in the microwave
range as the so-called ``negative bursts'' and also in the He~II
304~\AA\ line. The paper considers three eruptive events
associated with rather powerful flares. Parameters of absorbing
material of an eruption are estimated from multi-frequency records
of a ``negative burst'' in one event. ``Destruction'' of an
eruptive filament and its dispersion like a cloud over a huge area
observed as a giant depression of the 304~\AA\ line emission has
been revealed in a few events. One such event out of three ones
known to us is considered in this paper. Another event is a
possibility.
\end{abstract}

%\maketitle

\section{\textbf{Introduction}}

Prominences (filaments) in many solar events erupt and draw away
from the Sun as parts of Coronal Mass Ejections (CMEs). Partial
plasma flows from an eruptive prominence along its legs down to
the solar surface are often observed. ``Failed'' eruptions are
also known, when an eruptive prominence or filament after lift-off
rapidly decelerates and falls at nearly the same position where it
was located before the eruption \cite{Filippov2007}. Observations
of the 18.11.2003 eruptive event with the CORONAS-F/SPIRIT
telescope in the He~II 304~\AA\ line revealed one more scenario.
An eruptive filament disintegrated to form a Y-like cloud, flew
above the solar surface a distance of more than the solar radius
and, probably, landed far from the eruption site
\cite{Slemzin2004,Grechnev2005}. A similar anomalous eruption was
observed in the 13.07.2004 event \cite{Grechnev2008}. Detailed
observations of this event revealed an eruption from an active
region of a compact filament and its subsequent dispersion over a
huge area comparable with a quarter of the visible solar disk.
These events were accompanied with impulsive flares.

Eruptive events associated with flares produce various phenomena
of different temporal and spatial scales. These phenomena are
observed in diverse spectral domains (see \cite{Hudson2006}).
These are flare arcades whose emission is registered from X-rays
up to radio waves. Other phenomena are dimmings, i.e., depressions
of soft X-ray and extreme ultraviolet (EUV) emissions that reach
significant sizes and exist from a few hours up to two days. The
major cause of quasi-stationary dimmings is plasma density
decrease due to expansion of eruptive structures. Shock waves
excited by impulsive eruptions from active regions are manifest in
Moreton waves and some ``EUV waves'' (or ``EIT waves'').
Propagation of shock waves is traced from slowly drifting type II
bursts at metric and decimetric waves. Most likely, shock waves
form leading edges of fast CMEs, especially decelerating ones
\cite{Grechnev2010}.

Erupted plasmas with nearly chromospheric temperatures can occult
some part of the solar emission and absorb its detectable
fraction. Absorption phenomena are observed in various emission
ranges. Erupted plasmas show up in the H$\alpha$ line mainly as
surges, which are usually not large in size. In rare cases moving
clots can be detected far from an eruption site
\cite{Grechnev2008}. Observations of eruptions in the H$\alpha$
line are limited by the Doppler shift, which displaces them from
the filter band even if the line-of-sight velocities are rather
low. Other causes also rapidly decrease the optical thickness of
eruptions in the course of their expansion. Absorption phenomena
are also manifest in radio emission as the so-called ``negative
bursts'', i.e., a temporary decrease of the total radio flux below
the quasi-stationary level observed beyond bursts
\cite{Covington1953}. The optical thickness of eruptions in
microwaves is proportional to the wavelength squared. Unlike the
H$\alpha$ line, both static and eruptive filaments can be
optically thick practically in the whole microwave range, and they
cause appreciable depressions of the emission.

Absorption of the background solar emission due to
photo-ionization occurs also in the EUV range. Filaments are
sometimes observed as dark features in coronal emission EUV lines
Fe~IX-X 171~\AA\ and Fe~XII 195~\AA\ (often with a sheath
brightened during eruption), while dense parts of eruptive
filaments can appear as low-contrast moving fragmentary dimmings.
Absorption can be especially significant in the He~II 304~\AA\
line, because the cross section of photo-ionization by such an
emission of hydrogen-helium plasma with a nearly chromospheric
temperature is by an order of magnitude higher than that for the
195~\AA\ line. In addition, resonance scattering is possible in
the He~II 304~\AA\ line by plasma with a temperature of a few
$10^4$~K typical of the transition region (both between the corona
and chromosphere and between the corona and a prominence). The
temperature sensitivity maximum of the 304~\AA\ channel is
80\,000~K. Due to these properties, material of an eruptive
filament, whose temperature range can be sufficiently wide, might
have either increased or reduced brightness in this line. Finally,
erupted plasma can be also observed in the He~I 10830~\AA\ line,
which responds to both chromospheric and coronal phenomena. These
circumstances show that phenomena of long-lasting occultation of
the background solar emission can be observed in the He~II
304~\AA\ line and in radio emission in a range of 1\,--\,10 GHz
\cite{Kuzmenko2009}.

The paper \cite{Kuzmenko2009} analyzed eruptive events, for which
records of negative bursts at a number of frequencies were
available. Model estimations of temperatures, masses, and sizes of
absorbers from these data confirmed that the absorber in all the
cases was, most likely, material of eruptive filaments. The
present study addresses eruptive events followed by extended
darkenings in the SOHO/EIT 304~\AA\ channel and negative microwave
bursts. In Section~\ref{S-analysis} we analyze observations of the
events in various spectral ranges. In Section~\ref{S-estimate} we
estimate parameters of absorbing material for one of the events.
Sections~\ref{S-discussion} and \ref{S-conclusion} discuss and
summarize the results.

\section{Analysis of Observations}
\label{S-analysis}

For analysis of EUV observations we used images produced with the
EIT telescope \cite{Delab1995} on SOHO in the channels of 195~\AA\
(typical imaging interval of 12 min) and 171, 284, and 304~\AA\
(usually 6 hours). Source data FITS files were taken from the EIT
catalog\footnote{http:/\negthinspace/umbra.nascom.nasa.gov/eit/eit-catalog.html.}
We used total radio flux records made with Nobeyama Radio
Polarimeters\footnote{ftp:/\negthinspace/solar.nro.nao.ac.jp/pub/norp/xdr/},
RT-2 radio telescope of the Ussuriysk
Observatory\footnote{http:/\negthinspace/www.uafo.ru/observ\_rus.php},
and Learmonth station of the USAF RSTN Radio Solar Telescope
Network\footnote{ftp:/\negthinspace/ftp.ngdc.noaa.gov/STP/SOLAR\_DATA/SOLAR\_RADIO/RSTN\_1sec/}.
Data about CMEs were taken from the SOHO LASCO CME
Catalog\footnote{http:/\negthinspace/cdaw.gsfc.nasa.gov/CME\_list/;
\cite{Yashiro2004}}. Consolidated data on solar events were taken
from the Solar-Geophysical
Data\footnote{http:/\negthinspace/sgd.ngdc.noaa.gov/sgd/jsp/solarindex.jsp}.

\subsection{Event 1: 29.04.1998}

\begin{figure*}
  \centerline{
\includegraphics[width=\textwidth]{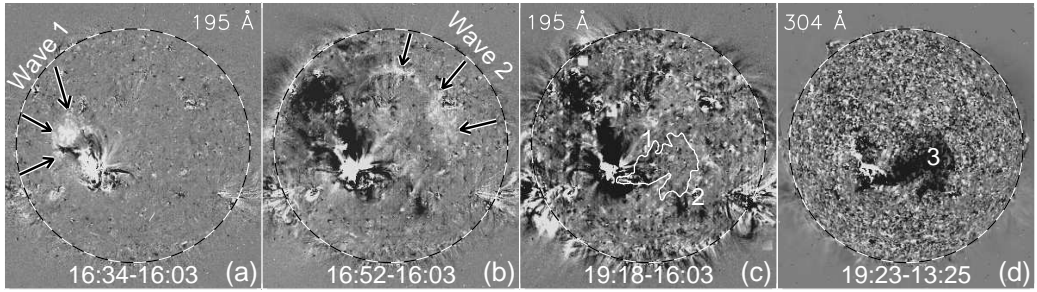}}
 \caption{Difference solar images for the
29.04.1998 event according to SOHO/EIT data.  (a\,--\,c): coronal
disturbances visible in the 195~\AA\ EIT channel; (d): extensive
darkening observed in the 304~\AA\ EIT channel. Black-white circle
denotes the solar limb.}
 \label{F-19980429_eit}
\end{figure*}

This event associated with a 3B/M6.8 flare (peak at 16:37, all
times hereafter are UT) occurred in active region NOAA~8210
(S16~E22, $\beta \gamma \delta$-configuration). Various aspects of
this events were addressed in a few papers
\cite{Wang2000,Chertok2003,Chertok2005}. Observations in the
H$\alpha$ line in the Big Bear Solar Observatory (BBSO)
\cite{Wang2000} and in the H$\alpha$ and He~I\footnote{see a movie
at
http:/\negthinspace/mlso.hao.ucar.edu/cgi-bin/mlso\_datasum.cgi?1998\&4\&29\&ACOS}
lines in the Mauna Loa Solar Observatory (MLSO) suggest that a few
sequential eruptions of different filaments or their parts
occurred in the event. The difference EIT 195~\AA\ and 304~\AA\
images in Fig.~\ref{F-19980429_eit}a\,--\,d show development of
disturbances caused by the eruptions. In all the images, the solar
rotation was compensated, and for each of the 195~\AA\ and
304~\AA\ channels, an image observed before the onset of the event
was subtracted (the processing technique is described in
\cite{Chertok2005BD}). First eruptions probably produced, at
least, two shock waves following each other that were manifest in
type II bursts and as EUV waves. The first EUV wave propagated in
the northeastern direction (Fig.~\ref{F-19980429_eit}a), and the
second one ran northwest (Fig.~\ref{F-19980429_eit}b). Material
erupted at the onset of the event was visible in the H$\alpha$
line at 16:32\,--\,16:46 to move in the NNE direction across the
equator. The brightest CME appeared above the northeastern part of
the limb at 16:58. The fast halo CME (average plane-of-sky speed
of 1374 km/s) significantly decelerated. Starting from 17:25 on,
an opposite part of the coronal transient was observed above the
southwest limb that was probably associated with a shock wave.

Large-scale quasi-stationary dimmings were registered in the
195~\AA\ channel (Fig.~\ref{F-19980429_eit}b,c). Three deepest
dimmings were almost invariable during 2 to 3 hours after the
eruption. One of them was located north from the eruption region,
the second one adjoined it from the south, and the third one was
located close to the northeastern part of the limb. Differences
between dimmings visible in the 171, 195, and 284~\AA\ lines were
not conspicuous. However, a single image observed during 6 hours
in the 304~\AA\ channel at 19:23 shows an extensive darkening west
from the eruption region without counterparts in the coronal
channels (Fig.~\ref{F-19980429_eit}d). Its area at a 25\% decrease
of the brightness reached 6.5\% of the solar disk, and the maximum
depth was $-85\%$ (the brightness drop to 15\% of an initial
level).

A number of observational facts indicates that darkening 3 was due
to occultation of the solar emission by erupted material. The
195~\AA\ images in an interval of 18:07\,--\,18:53 show a motion
of a long darkening 1 (Fig.~\ref{F-19980429_eit}c) towards the
extensive darkening 3. Changes are also visible related to a
southwest motion from the eruption region, towards darkening 2
(cf. Fig.~\ref{F-19980429_eit}b and \ref{F-19980429_eit}c). A slow
motion (less than 100 km/s) in the projection on the solar disk in
the NWW direction is detectable in the H$\alpha$ and He~I lines
after 16:52. This motion can be detected in the He~I images up to
20:00 at a distance from the eruption site almost equal to the
solar radius in the plane of the sky.

\begin{figure}
  \centerline{
\includegraphics[width=0.5\textwidth]{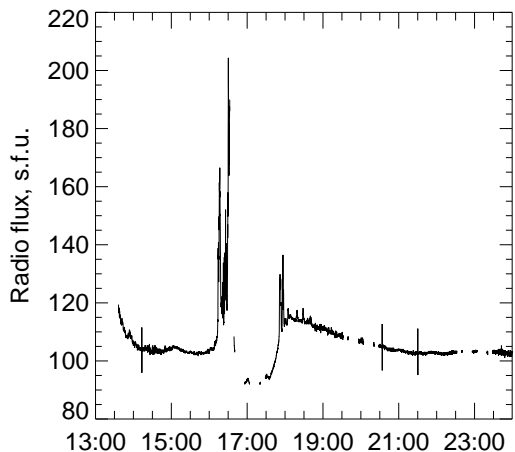}}
 \caption{ A record of the radio burst at a
frequency of 2.8 GHz made in the Pentincton observatory (Courtesy
K.~Tapping). }
 \label{F-19980429_neg_burst}
\end{figure}

A ``negative'' burst was observed in microwaves after an impulsive
burst. Fig.~\ref{F-19980429_neg_burst} presents a record made in
the Pentincton observatory at a frequency of 2.8~GHz [in solar
flux units, 1~s.f.u. = $10^{-22}$ W/(m$^2$~Hz)]. The record of the
burst was interrupted by calibrations, which take a significant
part of the observational daytime in this observatory (calibration
records were rejected in the figure to reveal the negative burst).
A decrease of the total radio flux relative to the
quasi-stationary level started by 16:56, lasted about one hour,
and reached $-11\%$ of the pre-burst level at the maximum
absorption (17:00\,--\,17:20). A negative radio burst as long as
53 min with a deepest depression at 16:40 was also recorded at 6.7
GHz in the Cuba observatory (Solar-Geophysical Data). A negative
burst in a range of 0.4\,--\,2.7 GHz is also surmised in
multi-frequency records of the Sagamore Hill and Palehua
observatories, but low quality of these records does not allow us
to quantitatively analyze them in order to estimate parameters of
an absorbing ``screen''. Nevertheless, by taking account of data
from different emission ranges it is clear that the decrease of
the total radio flux was caused not only by possible occultation
of microwave sources in an active region, but also by screening of
large quiet Sun's regions. No alternative to the eruption is seen
for the role of the screen.

These observational facts show that one or more filament eruptions
in the 29.04.1998 event were followed by disintegration of the
filament's magnetic structure and dispersing fragments over
significant area above the solar surface. They occulted the
background solar emission that caused the negative burst and the
long-lasting large-scale darkening visible in the 304~\AA\ channel
without pronounced analogs in coronal lines (although faint
manifestations of absorption were detectable in this region in the
195~\AA\ images). We do not exclude that the screens responsible
for the negative burst and the darkening in the 304~\AA\ channel
were formed by different anomalous eruptions. Obviously, an
increase of the microwave flux associated with a small radio
burst, which started at 17:50, must terminate the depression of
the total radio flux that might be longer if this burst would not
occur. However, the longer existence of the darkening in the
304~\AA\ line with respect to the negative burst (longer than
1.5\,--\,2 hours in this event) appears to be usual (see
Section~\ref{S-discussion}).

\subsection{Event 2: 27/28.05.2003}

Two eruptive events associated with big flares occurred in active
region NOAA~10365 (S07~W17, $\beta \gamma \delta$-configuration)
with an interval of about 90 min. The first event was associated
with a 2B/X1.3 flare (peak at 23:07 on 27 May), and the second one
was associated with a 1B/X3.6 flare (peak at 00:27 on 28 May).
Both events were observed in the H$\alpha$ line in
BBSO\footnote{http:/\negthinspace/bbso.njit.edu/} and in the
H$\alpha$ and He~I lines in MLSO\footnote{see a movie at
http:/\negthinspace/mlso.hao.ucar.edu/cgi-bin/mlso\_datasum.cgi?2003\&5\&27\&ACOS}.
The H$\alpha$ images suggest that Moreton waves were possible in
both events. Also, type II bursts were registered in each event.
These circumstances indicate excitation of shock waves in both
events that is consistent with a fact that the first halo CME
decelerated (average speed of 964 km/s). The SOHO/EIT telescope
observed at that time in the 304~\AA\ channel with a 12-min
interval, and images in all four channels were produced four times
per day as usually.

\begin{figure}
  \centerline{
\includegraphics[width=\textwidth]{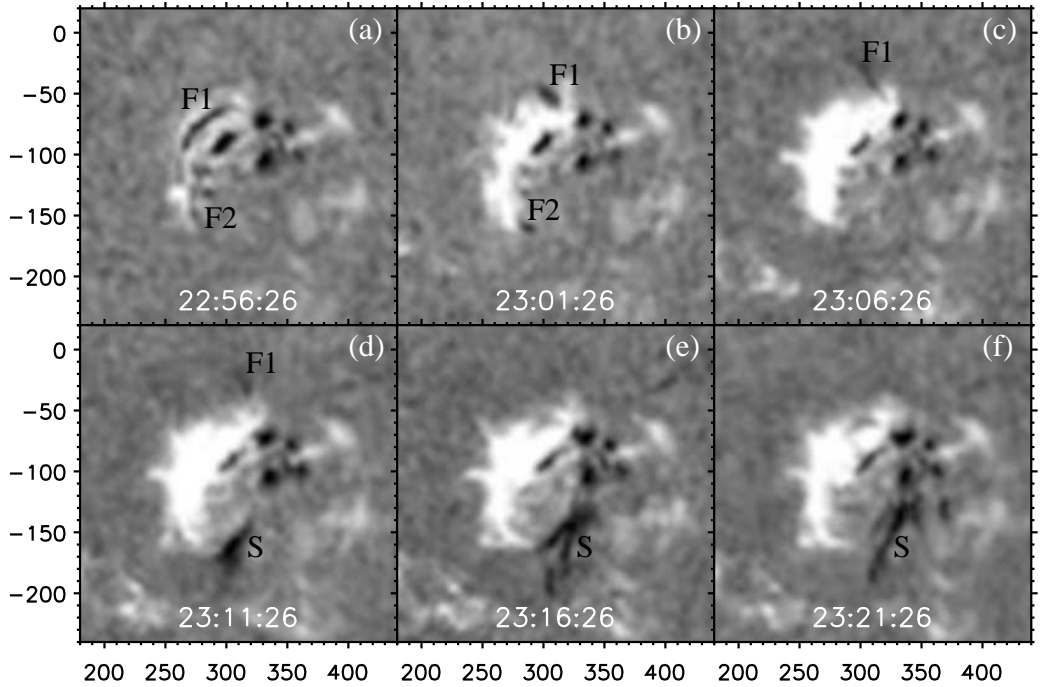}}
 \caption{The 27.05.2003 eruptive event in
H$\alpha$ images produced in BBSO. F1 and F2 are eruptive
filaments, S is a returning surge. Axes show hereafter arc seconds
from the solar disk center.}
 \label{F-20030527_h_alpha}
\end{figure}

Absorption phenomena were observed after the first event. The BBSO
H$\alpha$ images in Fig.~\ref{F-20030527_h_alpha} show that
filament F1 located in the northeastern part of the active region
started to erupt at about 23:00, and the southwards filament F2
disappeared (Fig.~\ref{F-20030527_h_alpha}b). The double eruption
is confirmed by the appearance of flare ribbons at the previous
locations of both filaments. A jet-like northern end of the
straightened filament F1 can be traced up to 23:11. A dark
returning surge appeared at that time in the southern part of the
active region presumably due to downflow of cool gas from an
expanding loop-like filament. This filament could be the F2
filament, but most likely this was a combined F1+F2 filament,
which formed during the eruption (see \cite{Uralov2002,
Grechnev2006}) .

\begin{figure*}
 \centerline{
\includegraphics[width=\textwidth]{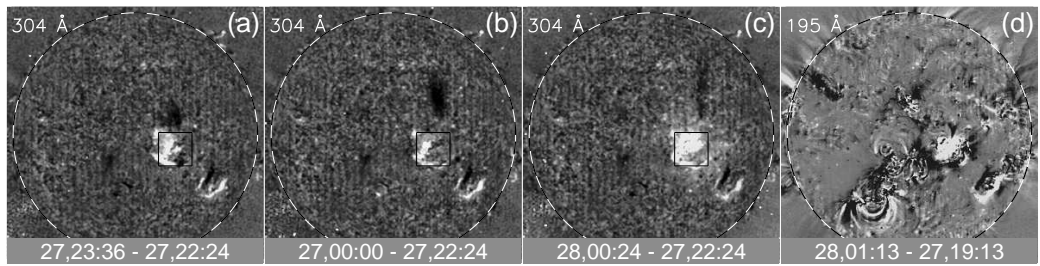}}
 \caption{Fixed-base difference images
compensated for the solar rotation produced from SOHO/EIT data
observed on 27/28.05.2003 in the 304~\AA\ (a\,--\,c) and 195~\AA\
(d) channels. The frame denotes the boundaries of the region shown
in Fig.~\ref{F-20030527_h_alpha}. The black-white dashed circle
denotes the solar limb. }
 \label{F-20030527_eit}
\end{figure*}

Five He~II 304~\AA\ images show an eruption, which looked like a
dark shadow moving in the NNW direction
(Fig.~\ref{F-20030527_eit}a\,--\,c), and appeared to be a
geometric extension of the straightened northern end of the
eruptive filament F1. The area of the ``shadow'' at a 25\% level
of the brightness decrease was about 1\% of the solar disk area,
and the depression depth reached $-65\%$. The eruption was well
visible also in the He~I 10830~\AA\ line. Dimmings observed
somewhat later in the coronal channels significantly differed in
positions, shapes, and sizes from the moving 304~\AA\ darkening
(Fig.~\ref{F-20030527_eit}). Thus, the northern darkening in this
event appears to have been a result of absorption of the
background solar emission in a dense part of a loop-like filament,
which expanded up to a huge size.

A halo, which appeared in SOHO/LASCO/C2 images at 23:50 or still
earlier, at 23:26, was presumably a result of impact of a shock
wave on coronal ray structures. Magnetoplasma structures appeared
in later images, but it is difficult to understand whether they
were ejected in the first event or in the second one. It is
possible that eruptive structures in both events continuously
expanded by keeping their integrity.

\begin{figure*}
\centerline{\hspace*{0.015\textwidth}
\includegraphics[width=0.515\textwidth,clip=]{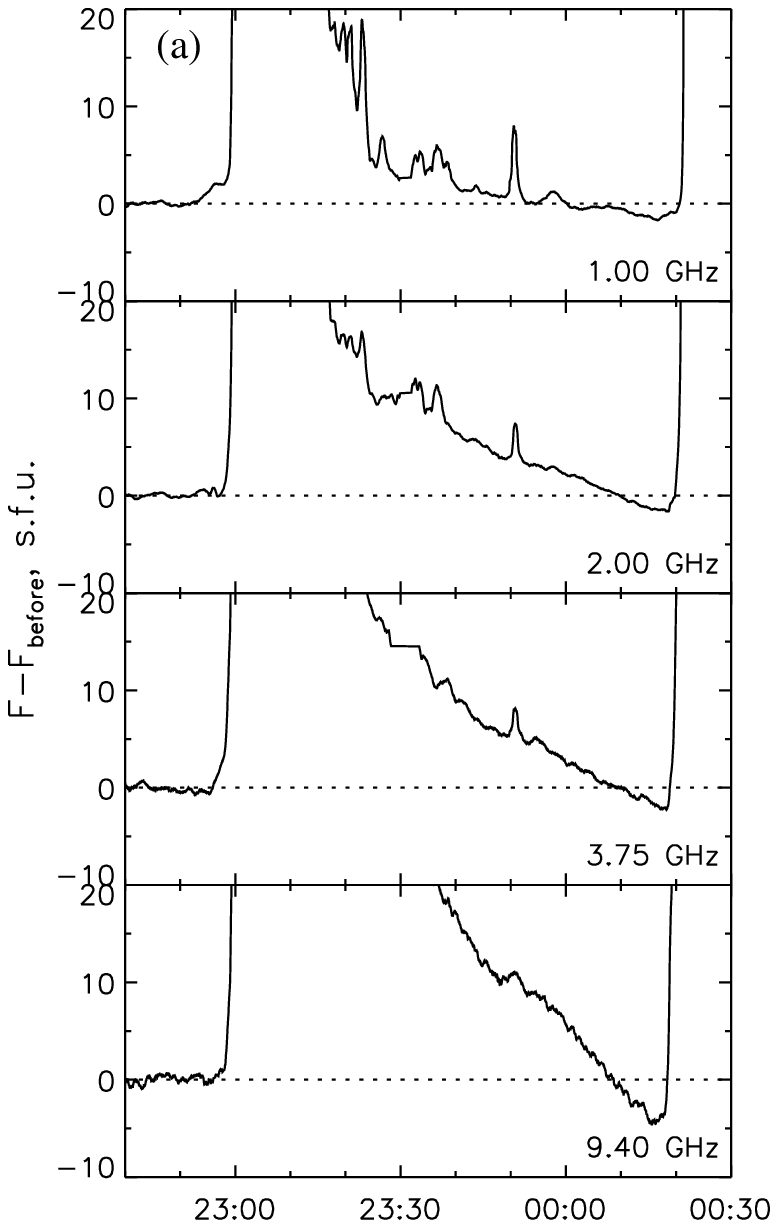}
% \hspace*{0.03\textwidth}
\includegraphics[width=0.515\textwidth,clip=]{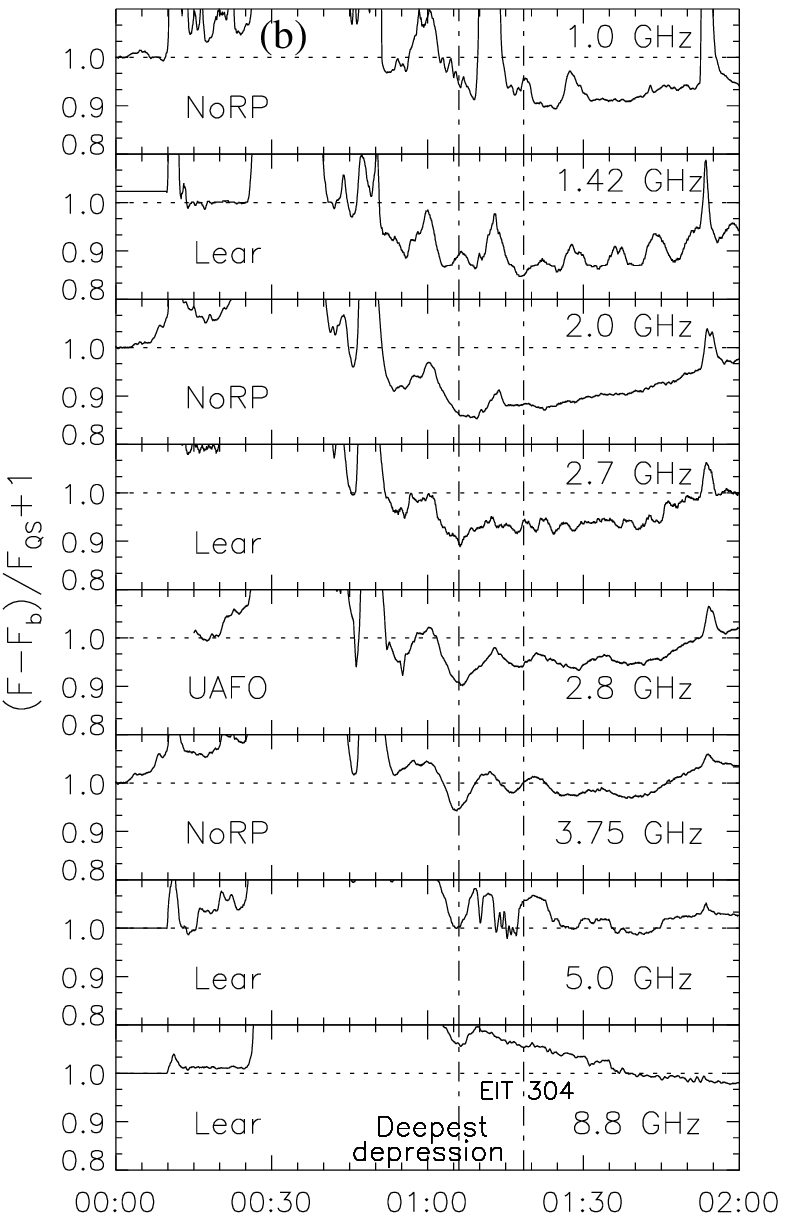}}
 \caption{Total radio flux time profiles at
different frequencies. a)~Background-subtracted data of Nobeyama
Radio Polarimeters recorded in the 27/28.05.2003 event. b)~Time
profiles of the 01.01.2005 event normalized to the quiet Sun's
level. The vertical dash-dotted lines in panel (b) mark the
instants of deepest depression (left) and the observation of the
darkening in the EIT 304~\AA\ channel (right). }
 \label{F-totflux_timeprof}
\end{figure*}

The flares in both events were accompanied by powerful microwave
bursts. Fig.~\ref{F-totflux_timeprof}a presents time profiles of
the first burst and the onset of the second one recorded with
Nobeyama Radio Polarimeters. To reveal a commenced negative burst,
the initial levels were subtracted, and fluxes were limited at a
level of 20 s.f.u. The decay of the first burst was followed by a
continuing flux decrease below the initial level, and then the
decrease was interrupted by the onset of the second burst. The
commenced depression is well pronounced, but it is not possible to
estimate parameters of the eruption from absorption of radio
emission.

     \subsection{Event 3: 01.01.2005}

\begin{figure}
  \centerline{
\includegraphics[width=\textwidth]{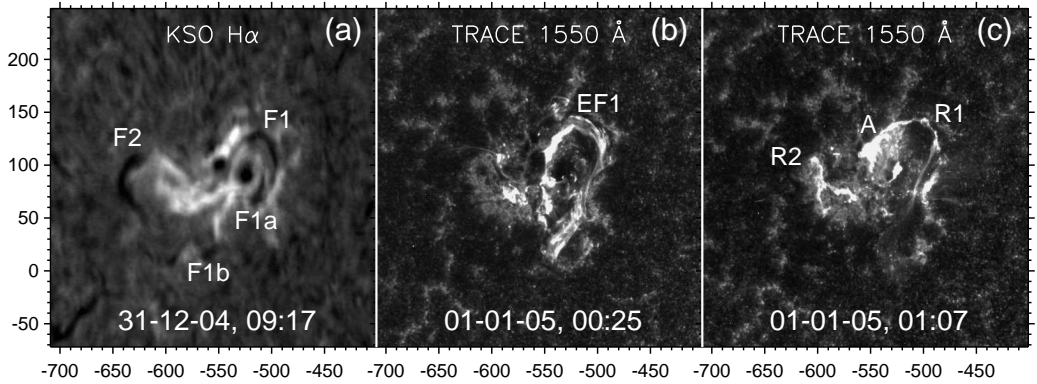}}
 \caption{Active region 10715 before eruption
on 01.01.2005 (a), after the C1.6 flare (b), and after the X1.7
flare (c).}
 \label{20050101_eruption}
\end{figure}

This event in active region NOAA~10715 (N04~E20, $\beta \gamma
\delta$-configuration) was associated with an X-ray flare of an
X1.7 importance (peak at 00:31) preceded by a weaker C1.6 flare at
00:13 in the same region. A type II radio burst and decelerating
CME (average speed of 832 km/s) were associated with the major
event. No core was visible in the CME structure. No H$\alpha$
patrol was carried out at that time, neither we found observations
of an eruption in this event. TRACE observed the first event
(C1.6) in UV channels and recorded jets in this active region. The
TRACE observations had a break from 00:26 till 01:06, when the
major event associated with the X1.7 flare occurred. The situation
before the event and after it is shown in
Fig.~\ref{20050101_eruption}. An H$\alpha$ image (Kanzelh{\" o}he
Observatory) in Fig.~\ref{20050101_eruption}a was obtained 15
hours before the event. In the active region there were filament
F1 near the eastern sunspot and one more filament F1a southward. A
southeast segment F1b was possibly located in the same filament
channel. The TRACE images observed in the C~IV 1550~\AA\ line (the
temperature sensitivity range of $(60-250)\times 10^3$~K) show the
situation after the C1.6 flare (Fig.~\ref{20050101_eruption}b) as
well as after the X1.7 flare (Fig.~\ref{20050101_eruption}c).
Fig.~\ref{20050101_eruption}b presents the onset of eruption of
the F1 filament (EF1), which might have combined with the F1a and
F1b filaments before the eruption or at its onset.
Fig.~\ref{20050101_eruption}c shows the decay phase of the X1.7
flare. A compact arcade A intruded into the eastern spot, and an
extension of the arcade northeast and then south, R1, was a flare
ribbon arranged along the former position of the F1 filament. A
brightening was visible in the western spot, too. Ribbons R2 were
observed in the eastern part of the active region, where the F2
filament presumably could intervene after the observations in
Kanzelh{\" o}he.

\begin{figure}
  \centerline{
\includegraphics[width=\textwidth]{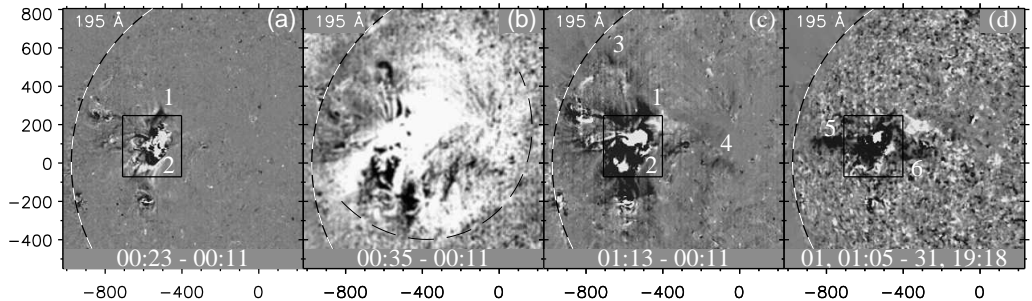}}
 \caption{ Fixed-base difference SOHO/EIT
images (solar-rotation-compensated) observed in the 01.01.2005
event in the 195~\AA\ (a\,--\,c) and 304~\AA\ (d) channels. The
frame denotes the boundaries of the region shown in
Fig.~\ref{20050101_eruption}. The black-white dashed circle
denotes the solar limb. The thin dashed arc in panel (b) outlines
the ``EIT wave'' front. }
 \label{20050101_eit}
\end{figure}

Difference solar images observed in the EIT 195~\AA\ channel in
Fig.~\ref{20050101_eit}a\,--\,c show the appearance and
development of quasi-stationary dimmings 1,~2 and  an ``EIT wave''
front outlined with the thin dashed arc in
Fig.~\ref{20050101_eit}b from the west and south. A low brightness
threshold was applied to reveal the wave front, therefore the
inner part of the image is excessively bright due to the flare
emission and stray light. The wave was also observed in GOES12/SXI
images in an interval of 00:32\,--\,00:37. A plane-of-sky wave
speed estimated from these images in the southern direction was
about 600 km/s. Short-lived moving darkenings, e.g., 3 and 4, were
also visible in the 195~\AA\ channel (Fig.~\ref{20050101_eit}c).

The 5\,--\,6 darkening in the 304~\AA\ channel
(Fig.~\ref{20050101_eit}d) strongly differs in the position and
shape from the major coronal dimmings 1 and 2 visible in the
195~\AA\ channel. The area of the 5\,--\,6 darkening at a level of
25\% brightness decrease was about 3\% of the solar disk area, and
a deepest depression reached $-70\%$. Like events 1 and 2, the
most probable cause of the darkening in the 304~\AA\ channel was
absorption of the background solar emission in cool material
ejected in eruption of filament F1 and probably F2. Significantly
weaker manifestations of absorption are also detectable in the
195~\AA\ channel that is consistent with the difference of the
absorption cross sections in these lines of an order of magnitude.

The darkening observed in the 304~\AA\ channel half an hour after
the eruption was much larger than filaments F1 and F2 before
eruption were. A CME at that time appeared in the LASCO/C2 field
of view, i.e., its distance from the eruption site exceeded the
solar radius. Occultation of the solar emission by an expanding
CME core was not excluded in the 27/28.05.2003 event, but it is
difficult to imagine such a situation for the 5\,--\,6 darkening
here. Like the 13.07.2004 event \cite{Grechnev2008}, the CME in
the 01.01.2005 event did not have a pronounced bright core, which
is usually identified with material of an eruptive filament. This
fact also agrees with an assumption that the main part of the
filament in the 01.01.2005 event was not ejected high into the
corona as a part of the CME, but returned to the solar surface.
Probably, a transformation of the CME magnetic structure occurred
during the eruption, and the main part of filament material had
not joined the CME.

A strong impulsive burst was observed in microwaves with a maximum
flux density of about 5000 s.f.u. at frequencies of 10\,--\,17
GHz. Following its decay, a negative burst occurred starting from
00:51\,--\,01:02 and up to 01:50\,--\,02:10.
Fig.~\ref{F-totflux_timeprof}b shows total flux time profiles of
radio emission recorded at a number of frequencies in
observatories of Nobeyama (1, 2, 3.75 GHz), Ussuriysk (2.8 GHz),
and Learmonth (1.4, 2.7, 4.9, and 8.8 GHz). Pre-burst flux levels
were subtracted, and the records were smoothed over 30~s and
normalized to the quiet Sun's level. The duration and depth of the
negative burst increased towards lower frequencies. The maximum
depth of $-13\%$ was observed at 01:06 at a frequency of 2~GHz.

\section{Estimation of parameters of absorbing material}
\label{S-estimate}

Possible causes of negative radio bursts have been addressed in
\cite{Sawyer1977}. Occultation of emission from a compact source
located in an active region by material of an eruptive filament
has been considered to be the major cause. The events which we
discussed as well as those analyzed in papers
\cite{{Grechnev2008}, {Kuzmenko2009}} show that absorption of
emission from compact sources is not the only factor. Absorption
of emission from significant areas of the quiet Sun can also
provide an important contribution. A dependence of absorption on
parameters of a screen and observing radio frequency provides an
opportunity to estimate these parameters from radio data recorded
at a number of frequencies. A model proposed in
\cite{{Grechnev2008}, {Kuzmenko2009}} allows one to calculate a
spectrum of the total solar radio flux by considering
contributions of a few layers. These are the chromosphere, a
screen (material of an eruptive filament) of a given area located
at a given height above the chromosphere, and coronal layers both
between the chromosphere and the screen, and between the screen
and an observer. By comparing an observed frequency distribution
of the depths of a negative burst with results of modeling, it is
possible to estimate the kinetic temperature of the screen, its
optical thickness at each frequency, its area, and its height
above the chromosphere.

\begin{figure}
  \centerline{
\includegraphics[width=0.5\textwidth]{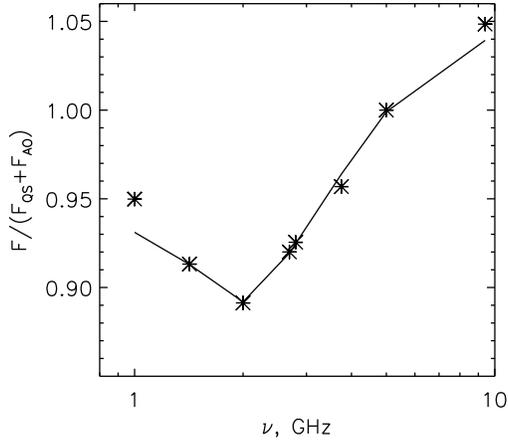}}
 \caption{The measured absorption depths
re-normalized to a sum of total radio fluxes from the Sun and an
occulted compact radio source (asterisks) and those calculated
with the model (line).}
 \label{20050101_spectrum}
\end{figure}

Multi-frequency records of a negative burst with a quality
sufficient to estimate parameters of an eruption are available for
event 3 (01.01.2005). Absorption depths at different frequencies
measured from data shown in Fig.~\ref{F-totflux_timeprof}b for the
instant of the maximum depth and re-normalized to the sum of total
radio fluxes from the Sun and an occulted compact radio source are
shown in Fig.~\ref{20050101_spectrum} with asterisks, and a
modeled spectrum is shown with the line. The depth of the negative
burst at 1~GHz could be reduced because of commencement of a type
III burst that explains the difference between the model and
observations at this frequency. The best fit of the observations
is achieved with the following parameters of an occulting screen:
the optical thickness at 2.8 GHz of 2.6, the kinetic temperature
of 14000~K, the effective height of the screen above the
chromosphere of 50 Mm, and the area of the screen about 5\% of the
solar disk area.

It is possible to estimate an average electron number density
$n_\mathrm{e}$ and mass of absorbing material $m$ by means of the
following expressions
%\begin{equation}
$$\tau=\frac{0.2n_\mathrm{e}^2L}{\nu^2T^{3/2}},\quad
m=m_\mathrm{p}n_\mathrm{e}SL,$$
% \nonumber
%\end{equation}
with $\nu$ being the radio frequency, $T$ the kinetic temperature,
$S$ and $L$ the area and geometrical depth of the screen, and
$m_\mathrm{p}$ being the proton mass. The ionization degree is
assumed to be about 100\%. The observations which we have
considered do not permit us to estimate the geometrical depth of
the screen. From an observed ratio of sizes of the 304~\AA\
darkening and assuming the geometrical depth of the absorbing
screen to be close to its width, we get $L\approx 70-100$~Mm, the
average electron number density $n_\mathrm{e} \sim
10^8$~cm$^{-3}$, and $m\sim10^{15}$~g, which is comparable with
masses of a typical filament and an ordinary CME. The estimate of
the mass is a highest one, but it nevertheless appears to be close
to a real value, because the dependence of the mass on the
geometrical depth is rather weak, $m \propto \sqrt{L}$.

The negative burst in event 1 (29.04.1998) was recorded with a
high quality at a single frequency of 2.8~GHz. By assuming the
temperature of the absorbing material of order 10000~K, and the
area of the occulting screen to be equal to the area of the
darkening observed at 304~\AA\ (6.5\%), the model allows us to
estimate the optical depth of the screen to be $\tau\approx
1.3-1.5$. With a maximum geometrical thickness of the screen
$L\approx \sqrt{S}\sim 300$~Mm, the average electron number
density $n_\mathrm{e}\sim 0.4\cdot10^8$~cm$^{-3}$, and the mass $m
\sim 2\cdot 10^{15}$~g that is of the same order as masses
estimated for a few events from multi-frequency radio data
\cite{Kuzmenko2009}.

\section{Discussion}
\label{S-discussion}

The three events considered in our paper as well as five events
analyzed previously in papers \cite{{Grechnev2008},
{Kuzmenko2009}} show that the most probable cause of depressions
of radio emission observed as ``negative bursts'' was absorption
of background solar emission in low-temperature plasmas of
eruptive filaments. This conclusion is in accordance with
estimated temperatures of occulting screens of $(8-14)\times
10^3$~K and their masses of $10^{15}$~g, close to values typical
of filaments. Just this interpretation of ``negative bursts'' was
initially proposed by Covington \cite{Covington1953} and confirmed
in most later studies \cite{Sawyer1977}. An essential result of
the analysis carried out in studies \cite{{Grechnev2008},
{Kuzmenko2009}} as well as in the present paper is the conclusion
about very large areas of occulting screens, from 2\% to 15\% of
the visible solar disk. It turns out that not only occultation of
compact sources in active regions is important, but also
occultation of large quiet Sun's areas. Huge sizes of occulting
screens are in agreement with a maximum thickness of about 200~Mm
estimated from observations of one of such events
\cite{Grechnev2008} as well as heights of their lower edges of
20\,--\,80~Mm above the chromosphere estimated by means of the
model.

In order to produce an appreciable depression of radio emission,
an eruptive filament must have a sufficient contrast against the
background of the quiet Sun with a brightness temperature of
$T_\mathrm{QS}$, while the corona must be optically thin. These
conditions are mainly satisfied within a frequency range from
1~GHz ($T_\mathrm{QS}=153000$~K) to 10~GHz
($T_\mathrm{QS}=12000$~K), in which negative bursts in mentioned
studies have been actually registered. Observations of negative
bursts beyond this frequency range appear to be possible in some
cases. However, with the estimated temperatures of screens of
$(8-14)\times 10^3$~K, frequencies below 5~GHz
($T_\mathrm{QS}=18000$~K) are favorable. Indeed, depressions in
all eight cases were deepest at frequencies of 2\,--\,4~GHz, and
the fact that Covington discovered negative bursts in observations
at a traditional wave of 10.7~cm (2.8 GHz) does not seem to be
surprising.

As noticed in Introduction, the He~II 304~\AA\ line of the EUV
range is another spectral domain, which favors detection of
large-scale ejections of low-temperature plasmas of eruptive
filaments due to absorption. This is due to a high cross section
of photo-ionization by this emission and the existence of an
additional mechanism, resonance scattering of emission by plasmas
with temperatures of a few $10^3$~K. For these reasons,
depressions caused by absorption of the He~II 304~\AA\ line
emission are not visible or scarcely detectable in coronal
emission lines, unlike deep dimmings caused by plasma density
decrease due to a CME expansion. It is therefore clear that
observations of such darkenings in the He~II 304~\AA\ line
correlate with negative radio bursts. Note that the depressions
observed in the He~II 304~\AA\ line in the considered events
1\,--\,3 were significantly deeper than the 50\%-level expected
for resonance scattering (see \cite{Grechnev2008}) that confirms
predominance of low temperatures in occulting screens.

However, significant differences exist between absorption
properties in the microwave range, on the one hand, and in the
He~II 304~\AA\ line, on the other hand. The major difference is
determined by dissimilar density dependence of the optical
thickness of an absorbing screen. The microwave optical thickness,
$\tau \sim n_\mathrm{e}^2L \sim L^{-5}$, rapidly decreases in
omnidirectional expansion of a screen, if the number of particles
responsible for absorption remains constant in the screen. The
situation is different when dealing with absorption in the He~II
line. By neglecting temperature effects and other ones, on can
consider in the first approximation the optical thickness to
depend on the first power of density, $\tau \sim n_\mathrm{e}\,L
\sim L^{-2}$, and the role of the expansion factor is
significantly weaker. Even if a screen expands dissimilarly in
different directions, the density decrease factor dominates in the
radio range.

The second distinction is due to differences between layers
responsible for emissions registered in the 304~\AA\ channel, on
the one hand, and in microwaves, on the other hand. The background
solar emission in the 304~\AA\ line is most likely dominated by
contribution from the transition region, i.e., at very low
altitudes ($<10$~Mm). The contribution from coronal layers is
presumably dominated by the Si~XI line emission. It does not seem
to be significant in quiet regions. The situation is different in
the radio range. The quiet Sun's brightness temperature observed
from the Earth consists of two components, the chromospheric
emission ($\approx 10000$~K) and the coronal one. Unlike the
304~\AA\ channel, the corona contribution in microwaves exceeds
the chromospheric one at frequencies below 4.4~GHz. This
contribution gradually increases with a rise from the chromosphere
upwards. Therefore, the height of an absorbing screen above the
chromosphere is very important. For example, the presence at a
large height of an opaque, dense (for simplicity) screen with a
temperature of 14000~K decreases a low-frequency total radio flux
and increases a high-frequency one. By contrast, a very low screen
with the same parameters but the temperature of 10000~K would not
manifest in radio emission at all. However, a darkening in the
304~\AA\ channel is expected to be the same in both cases. One
might expect for these reasons that a negative burst should
typically terminate earlier than a darkening in the 304~\AA\
channel disappears. The latter might be visible as long as an
absorbing screen is located higher than the transition region.

Thus, phenomena of long-lasting screening of the background solar
emission can be observed in the He~II 304~\AA\ line as well as in
radio emission of the 1\,--\,10~GHz range (preferably below
5~GHz). However, SOHO/EIT supplies, as a rule, one 304~\AA\ image
in six hours, and therefore the probability to register such a
phenomenon is low. STEREO and SDO space-borne observatories carry
out observations in the 304~\AA\ channels in more detail, but they
operate during a prolonged minimum of the solar activity. Out of
very few heliographs, the SSRT \cite{Grechnev2003} is the only
one, which routinely observes the Sun in the frequency range of
interest during the whole light day. However, screening phenomena
after flare-related eruptions have not been detected in its
observations so far. Probably, this is due to both infrequent
occurrence of such events and an insufficient sensitivity of this
instrument that is required for their detection at its rather high
frequency of 5.7~GHz ($T_\mathrm{QS}=16000$~K) in the presence of
bright flare emissions. Total radio flux records persist to be the
main source about such events, but their detection in such records
is obviously preconditioned by a short duration of a preceding
flare and the absence of bursts afterwards. For all of these
reasons, information about such events is very limited.

The brightness decrease of solar disk portions below the
quasi-stationary level in the discussed events was caused by
absorption of a fraction of the solar emission in a cool screen.
Plasmas of eruptive filaments (prominences) constitute such
screens. Two scenarios of screening are possible. In the first
case, the shape and magnetic structure of an eruptive filament do
not undergo essential changes. Such a filament draws away from the
solar surface almost without loss of its mass and looks like a
moving screen, whose sizes increase, while the opacity decreases.
Our event 2 (27/28.05.2003) is a possible example of such a
scenario. This scenario is typical of non-flare-related filament
eruptions outside of active regions. Many such events have been
observed with SSRT and the Nobeyama Radioheliograph (frequencies
of 17 and 34 GHz). In the second scenario, the whole mass of an
eruptive filament or its considerable portion does not leave the
Sun as a part of a CME, but is dispersed over a large area above
the solar surface. This presumably occurred in event 1
(29.04.1998), and also in the 13.07.2004 event
\cite{Grechnev2008}. This scenario is not excluded also for event
3 (01.01.2005). Dispersion occurs along magnetic field lines,
which probably were not connected with the filament before the
eruption. Transposition of cool plasmas of the eruptive filament
to the external field lines is possible in reconnection of
magnetic fluxes of the internal field belonging to the filament
and the external one. One may speak in this case about
``destruction'' of the magnetic structure of the eruptive
filament. Afterwards, the ejecta does not resemble itself as it at
the onset of the eruption was. An example is the 18.11.2003 event
mentioned above \cite{{Slemzin2004}, {Grechnev2005}}, in which an
eruptive filament encountered an invisible obstacle and visually
transformed into two wide diverging jets carrying material of the
filament to remote sites on the solar surface.

It is a subject of a separate analysis how ``destruction'' of the
magnetic structure of an eruptive filament occurs. Now we only
indicate possible causes. The first one has been confirmed by
observations. Its essence is a pass of an eruptive filament
through a null point of the magnetic field in the corona (see,
e.g., \cite{Gary2004}). Indications at essentially
three-dimensional nature of an occurring transformation of an
eruptive filament into eruptive jet can be found in papers
\cite{{Meshalkina2009}, {Filippov2009}}. Another reason of allied
nature might be a pass of an eruptive filament across vicinities
of a point, in which loci of neutral lines of the radial magnetic
component bifurcate. Such a region can be revealed from the
arrangement of neutral lines of the large-scale magnetic field
calculated at different heights above the solar surface. A rising
eruptive filament is known to successively reproduce in its motion
contours of the same neutral line
\cite{{Filippov2001},{Filippov2002},{Filippov2008}} changing with
height. This lines breaks in a bifurcation point, and the filament
has no chance to retain its integrity after passing through such a
region. We found a similar bifurcation of the height distribution
of neutral lines in the mentioned 18.11.2003 event, which will be
analyzed in a separate paper.

An anomalous eruption with a ``destruction'' of the filament's
magnetic structure is favored by the appearance in real solar
conditions of null points and bifurcation points on the way of a
filament. This is more probable in a complex coronal magnetic
configuration, inside which the eruption occurs. Outer
manifestations of such a situation are complexity of a
configuration observed on the photosphere (all the three
considered events occurred in $\beta\gamma\delta$-configurations)
and/or surrounding of the eruption region by other active regions
(this was the case for events considered in \cite{Kuzmenko2009}).
These circumstances correspond to a conclusion of Sawyer
(\cite{Sawyer1977}, item 4 in `Summary and Conclusions') about a
tendency of negative bursts to clustering in complexes of active
regions. For the same reasons, an anomalous eruption of a
quiescent filament beyond an activity complex is not expected.
After eruptions occurring outside of active regions, one might
expect occultation of the solar emission by an expanding filament,
but its ``destruction'' is unlikely.

All the considered events were indeed associated with filament
eruptions from active regions accompanied by pronouncedly
impulsive flares. This is also supported by the appearance of
coronal shock waves in the events. Observations suggest that
excitation of a shock wave in the low corona is most probable by
the impulsive-piston mechanism in an eruption of a magnetic flux
rope structure \cite{Grechnev2010}. Then the shock wave detaches
from the piston, which excited it, runs ahead, and freely
propagates afterwards like a blast wave. The front of such a wave
has a shape close to a spheroid. Near-surface signatures of such
shock waves can be Moreton waves and those ``EIT waves'', which
propagate far away from an eruption region. Signatures of shock
waves propagating away from the solar surface are metric type II
bursts and outer edges of coronal transients of a halo type.
Kinematical correspondence of all of these manifestations has been
shown in the same study \cite{Grechnev2010} as well as a low
probability of other discussed excitation mechanisms of shock wave
excitation near the Sun.

Observations of all the three considered events have revealed type
II radio bursts in the metric range, ``EIT waves'' propagating up
to large distance from eruption regions, and indications of a
Moreton wave in event 2. Sufficiently fast (average speeds
$>830$~km/s) halo coronal transients were associated with all the
events, and all of them decelerated. An additional hint at a wave
character of these transients is a significant difference between
the expansion directions of their fastest parts and the major CME
magnetoplasma structures. A trace of a spheroid front of a coronal
shock wave is expected to be observed just in this manner.

\section{Conclusion}
 \label{S-conclusion}

We have demonstrated manifestations in different spectral ranges
of eruptive events associated with powerful flares followed by
long-lasting screening of the solar emission. Absorption of the
background emission in material of an eruptive filament can be
observed as a depression of the He~II 304~\AA\ line emission
without pronounced counterparts in EUV emission lines. It can be
also observed as a negative radio burst in the microwave range.
Darkenings in images obtained in the He~II 304~\AA\ channel can be
detectable well after the end of a corresponding negative radio
burst. From high-quality multi-frequency microwave data by means
of the model developed, it is possible to estimate parameters
(including mass) of ejected material, whose projection is on the
solar disk. Such estimates have confirmed that screening occurred
by material of an eruptive filament and showed that, besides
occultation of emission from compact sources in active regions,
occultation of significant quiet Sun's areas is significant.

A steadily expanding eruptive filament is known to be able to
occult the solar emission during tens of minutes or even a few
hours. Observations show that in some cases, which are presumably
rare, an eruption scenario can be anomalous. That is, an eruptive
filament encounters an unavoidable obstacle in a form of a
peculiarity of the coronal magnetic configuration and
disintegrates into parts or a cloud of fragments. In such a case,
material of an eruptive filament does not join a CME, but instead
is dispersed over a large area above the solar surface and
eventually lands far from the eruption region. Being aware of only
a few examples of such events, it is difficult to judge about
their common properties. We can only present some conjectures.

An anomalous eruption is expected to be favored by complexities of
a magnetic configuration, especially of the $\delta$ type, and
surrounding of the active region with others. Such an eruption
might be accompanied by a significantly power flare, surges, or
sprays. Surges observed in the H$\alpha$ line might sometimes
present only the slowest and densest part of an eruption, whose
real sizes can be significantly larger. In an anomalous eruption,
appearance of a shock wave is highly probable, which can manifest
in a metric type II burst, ``EUV wave'' propagating far from the
eruption region, and a possible Moreton wave. The shock wave might
be responsible for the leading edge of a coronal transient
originating in such an event. The CME can be without a pronounced
core. The outer CME edge might be either formed by a shock-driven
plasma flow or constituted by coronal rays deflected by the wave.
Such a coronal transient probably has a high speed and
decelerates. A number of listed properties appears to be typical
of many flare-related eruptions. If such an event is followed by a
negative radio burst or a large darkening observed in the 304~\AA\
channel that conspicuously mismatches dimmings visible in coronal
lines, then an anomalous eruption might have occurred in this
event.

The authors thank V.~A.~Slemzin for useful discussions and
K.~Tapping for data of the Pentincton Observatory. We are grateful
to team members of the Nobeyama and Learmonth observatories for
the opportunity to use their total flux radio data at different
frequencies. We thank the SOHO consortium for data used in the
analysis (SOHO is a project of international cooperation between
ESA and NASA). We used data from the CME catalog generated and
maintained at the CDAW Data Center by NASA and the Catholic
University of America in cooperation with the Naval Research
Laboratory. The research was supported by the Russian Foundation
of Basic Research (grants 09--02--00115, 11--02--00038, and
11--02--00050), Integration Project of RAS SD No.~4, and the
programs of basic researches of RAS ``Plasma Heliophysics'' and
``Solar Activity and Physical Processes in the Sun\,--\,Earth
System''.

\newpage

\end{document}